\documentclass[pre,floatfix,amsmath,amssymb,showpacs]{revtex4}
\usepackage{graphicx,color}
\usepackage{dcolumn}
\usepackage{bm}

\begin{document}

\input psfig.sty

\title{Star polymers: A study of the structural arrest in presence of attractive interactions}

\author{F. Lo Verso $^{1}$, L. Reatto $^{1}$, G. Foffi $^2$, P. Tartaglia $^2$, K. A. Dawson $^3$}

\affiliation{ $^1$ Istituto Nazionale di Fisica della Materia
and Dipartimento di Fisica,
Universit\`a di Milano, Via Celoria 16, I-20133 Milano, Italy\\
$^2$
Dipartimento di Fisica and INFM
Center for Statistical Mechanics and Complexity,
Universit\`{a} di Roma La Sapienza, Piazzale Aldo Moro 2,
I-00185 Rome, Italy\\
$^3$
University College Dublin,
Irish Center for Colloid Science and Biomaterials,
Department of Chemistry, Belfield, Dublin 4, Ireland}

\date{\today}

\begin{abstract}

\noindent Simulations and Mode-Coupling Theory calculations, for a large range of   the arm number $f$ and
packing fraction $\eta$ have shown that the structural arrest and the dynamics of star
polymers
 in a good solvent
 are extremely rich:
the systems show a reentrant melting of the disordered glass nested between two stable
fluid phases that strongly resemble the equilibrium phase diagram. Starting from a simple
model potential we investigate the effect of the interplay between attractive interactions
of different range and ultrasoft core repulsion, on the dynamics and on the occurrence of
the ideal glass transition line.  In the two cases considered so far, we observed some
significant differences with respect to the purely repulsive pair interaction.  We also
discuss the interplay between equilibrium and non equilibrium phase behavior. 
The 
accuracy of the theoretical tools we utilized in our investigation has been
checked by comparing the
results with molecular dynamics simulations.

\vskip 0.55truecm

\noindent
E-mail: federica.loverso@mi.infn.it
\end{abstract}
\maketitle

\section{Introduction}
\label{INTRO-sec} When a substance dissolves into another to form
 a true solution and the units dispersed through the solvent are much
 larger in size than the solvent molecules, we name it a {\it
 colloidal dispersion}. There are different classes of colloidal
 suspensions: molecules individually larger than $1 nm$\cite{footnote}
 e.g. proteins, polysaccharides, polymers, as well as dispersions that
 arise when a number of small molecules associate together to form an
 aggregate, it is known, for example, that amphiphilic molecules in a
 suitable solvent, above a critical concentration, can aggregate in
 micelles \cite{librorosa}. Varying the interactions between the
 mesoscopic constituent particles in colloidal dispersions results in
 a broad range of equilibrium and non-equilibrium fluid behavior.

\noindent 
One strategy to study such systems involves summing over the solvent
degrees of freedom, leaving an effective interaction between the
center of mass of the macroparticles \cite{Pusey}. The resulting
effective potential $V(r)$ depends, in a complicated fashion, on the
interparticle separation and it is in general state dependent
\cite{dendrimeri2}, \cite{Jpierre}. A special example in this context is given by the pair
potential describing a star polymer solution.

Star polymers (SP) can be considered as a generalization of
polymer-coated colloidal particles in the limit where the number of
monomers (N)  per chain is large and the size of the central core is
small with respect to the whole star extension.  The repulsive
interaction between star polymers at short distances increases very
slowly as the interparticle separation decreases, namely in a
logarithmic way. The number of chains chemically linked to a core
influences the `softness' of $V(r)$: stars with small arm number $f$
may interpenetrate widely; in the limit case $f=1,2$ the star polymer
reduces to a simple polymer chain. Stars with very large $f$ emphasize
their colloidal nature, although only in the limit $f\to \infty$ the
potential resembles the hard-sphere one.

\noindent Recently solutions of star polymers have received attention in relation to
several medical and industrial applications \cite{industrial}.
Moreover, in the last ten years, advances in macromolecular chemistry,
leading to the synthesis of regular star polymers, have made it
possible to explore the physics of very different model systems
monodisperse in {\it N} and {\it f} \cite{td65}, \cite{harreis10}.
Finally these systems are very interesting from a theoretical point of
view in relation to their polymer-colloid hybrid character. Indeed the
efficient design of new mesoscopic materials with properties
intermediate between different classes of colloidal systems is a very
important challenge in soft condensed matter physics.

In the last decades the equilibrium and non-equilibrium phase behavior of SP solutions have
been widely investigated, both theoretically \cite{td47,anom,Phtd,SCH2} and experimentally
\cite{td67,td53,industrial,td49}. These system are an interesting example of a
complex fluid for which the phase diagram has special features arising from the ultrasoft
nature of the repulsive interaction: e.g. there exists a cutoff value of the functionality
$f$ below which the system is fluid for all densities $({f}_c=34)$ and for $f>{f}_c$ the
phase diagram exhibits several unusual solid lattices as well as reentrant melting
\cite{Phtd}; SANS and SAXS experiments on solutions of many-armed stars above $f_c$, e.g.
\cite{td118},
 have revealed
different macrocrystal structures as we  increase the density. The functionality-dependent
bcc- and fcc-solids \cite{gastprl93,gastpre96}, as well as the reentrant  melting
transition \cite{gastmacrom97}, have been experimentally observed in solutions of
star-like block copolymer micelles. A second freezing transition observed in the same
experiments can be interpreted as the freezing in a bco crystal \cite{ref1}.

The dynamical properties of star polymers have been extensively
investigated.  Several studies, focused on star-polymers or star-like
systems in athermal solvent with different arm numbers, have shown
that it is quite difficult to nucleate a crystal: in many cases,
mainly at high functionalities, the solutions display a gelation
transition
\cite{15,vlassopoulos:jpcm:01,kapnistos:prl:00,loppinet:macrom:01,stiakakis:prl:02,stiakakis:pre:02}.
Molecular Dynamic (MD) and Brownian Dynamic (BD) simulation data for a
large range of $f$ and packing fraction $\eta=\frac{\pi}{6}\rho
{\sigma}^3$ show that the dynamics of star polymer solutions is
extremely rich.  In particular the ideal glass transition line,
obtained by means of the Mode-Coupling Theory (MCT) \cite{mct},
displays a non-monotonic behavior as a function of $\eta$ and
$f$. This behavior has been connected to the $\eta$ and $f$-dependence
of the effective hard core diameter of an equivalent HS system.  The
detailed comparison between theoretical predictions and simulations
confirms the validity of the MCT approach to study the disordered
arrested states in soft matter like colloids \cite{fsnat,ken} and in
ultrasoft systems like star polymer solutions \cite{artvetri}.  In
particular it has been confirmed that the modified hypernetted chain
integral equation (MHNC) is a very good approximation to study static
correlations in systems described by ultrasoft interactions not only
in equilibrium \cite{artmio} but also in metastable states
\cite{artvetri}.

Recently a model potential has been suggested to describe $SP$
solutions where, in addition to the excluded volume effects, attraction
emerges due to dispersion or depletion forces.
For this model the fluid-fluid phase diagram has been determined \cite{artmio}
using
mean field theory and two fluid-state-theories, MHNC and the
hierarchical reference theory (HRT) \cite{H3,H4,softHRT} for different $f$.
This analysis shows that
when the strength of the interaction is strong enough a
fluid-fluid phase transition appears but the density-temperature
coexistence curve bifurcates at a triple point into two lines of
coexistence terminating at two critical points.  This peculiar phase
behavior is related to the unusual form of the repulsive
contribution.

\noindent
Moreover it has been shown that
self-organized structures, resulting from telechelic linear homopolymers and copolymers, similar to star polymers, can bridge by producing
an effective attractive interaction leading
 to reversible aggregation of macromolecules \cite{semenov}.
These micelles are constituted by telechelic associative
polymers which  have the associating groups at the chain ends.
Above a critical  concentration
the end groups associate in multiplets, forming
flower-like polymeric micelles.
 At higher concentrations the process of bridging can lead to the formation
of a transient gel or also induce macroscopic phase
separation \cite{taco0,taco}.

The  aim of this paper is to investigate  the effect of attractive interactions on the dynamics and on the occurrence of the ideal glass transition line,
trying to emphasize
the special features of the phase behavior
arising from the ultrasoft nature of this repulsive
effective
interaction.
We
also qualitatively discuss  the interplay between equilibrium and non equilibrium
phase behavior.
In this work different theoretical and numerical  methods
have been utilized
(mean field theory, fluid state theories, mode-coupling theory, molecular
dynamic simulations).

\noindent We recall that in the last years a great number of studies focused on the
dynamical behavior of short ranged attractive systems, which are characterized by a
strong repulsive core besides the attraction. In particular when the range of the
attraction becomes much shorter than the typical diameter of the colloids, phenomena like
a reentrant glass transition or the existence of two different glassy phases emerge
\cite{Fabbian1999,Bergenholtz1999,Dawson2001}. This peculiar behavior has been confirmed
by a large number of simulations \cite{Puertas2002,Zaccarelli2002b,Puertas2003,foffi} and
experiments \cite{Mallamace2000,Pham2002,Eckert2002}. Notice  that, historically, these new
findings have been predicted for the first time within MCT calculation and only on a
second stage confirmed by experiments.  Consequently it is clear that this kind of
approach can be extremely useful also for different interaction models. It is interesting
now to focus the attention on ultrasoft repulsion (typical for example of a SP
solution) and an attraction with different range (which could be typical, for example, of
depletion interactions) and to investigate the possibility of new features.

This paper is organized as follows: in section \ref{METHsec} we
present the general framework for our research.  In particular, in
subsection \ref{INTE-sub} we introduce the interaction model we chose
to study star polymers in presence of attraction.  Then in subsections
\ref{MHNC-sub}, \ref{MD-sub} and \ref{MCT-sub} we describe the
theoretical and numerical tools we used
to study the structural properties, the structural arrest and the
diffusivity in dense star polymer solutions: MHNC, MCT and MD,
respectively.  In particular we discuss the application of the
modified hypernetted chain integral equation to study the structural
properties of star polymer solutions and some test of its accuracy
when attractive interactions are taken into account in addition to the
entropic contribution.

\noindent In section \ref{RESULTS-sec} we  consider the effect of attraction 
on the slow dynamics and structural arrest of star polymer
solutions. We carried out a mode-coupling theory analysis which allows
us to locate the nonergodicity transition curve of the system, using
as input the information on the structure obtained by MHNC.  Our aim
is to complete the picture of the phase diagram of a star polymer
solution in the presence of attractive forces, investigating the
dynamics, for the values of the parameters which govern the intensity
of the attraction extensively discussed in
Ref.~\cite{artmio}.  Then we modify these parameters in such a way to consider attractive
forces of shorter range, and we focus on the effect of  these modifications on the
properties of the glass state. In order to test this difference, molecular dynamics
simulations have been performed and the diffusivity of SP fluid as been evaluated up to
crystallization. In this section we again discuss the structure of the system very close
to the glass phase.

\noindent
Finally in section \ref{Concl-sec} we  discuss and summarize our results
and we  draw our conclusions.

\section {Introduction to the model and methods of study\label{METHsec}}
\subsection{Effective pair interaction}
\label{INTE-sub}
The effective pair interaction between star polymers with $f$ arms
in a good solvent is
purely  repulsive and for $f\geq 10$ it reads  as follows:

\[
\frac{V_{rep}(r)}{k_{B}T}=\frac{5}{18}~f^{\frac{3}{2}}\left[-\ln(\frac{r}{\sigma})~+
(1+\frac{\sqrt{f}}{2})^{-1}\right]~~~~~~~~~~~~~~~~~~~~~(r\leq\sigma)
\]

\begin{equation}\label{Vot2}
~~~~~~~~~~~~~=\frac{5}{18}~f^{\frac{3}{2}}(1+\frac{\sqrt{f}}{2})^{-1}
(\frac{\sigma}{r})\exp\left[\frac{-\sqrt{f}(r-\sigma)}
{2\sigma}\right]~~~~~~~~~~(r>{\sigma})
\end{equation}

\noindent
$\sigma$ is the corona diameter of the star and  depends on the number $N$ of monomers of a single arm
\cite{Da}.
Witten and Pincus determined by scaling theory the explicit form of such interaction
at short distances \cite{td47};
 it has been also shown that
the good agreement between theory and experiments significantly
improves if the model interaction has, in addition to a logarithmic repulsive core,
  a long range interaction of
Yukawa  form
\cite{anom}, \cite{td49}, \cite{td50}, \cite{td126}.
This model interaction gives a good description of small angle neutron scattering (SANS)
 results on concentrated star polymer samples \cite{td49} for an wide range of $f$ values.
For $f\leq 10$   a logarithmic-Gaussian potential more accurately describes the effective interaction  \cite{low2001}
but we do not study this regime.

In a previous paper \cite{artmio}
the phase diagram of star polymer solutions
has been investigated
when
the effective
interaction between two star polymers contains an additional attractive contribution
$w(r)$ to be added to the repulsive part $V_{rep}(r)$ of equations (\ref{Vot2}):

\begin{equation}\label{VW}
V_{tot}(r)= V_{rep}(r) + w(r).
\end{equation}

\noindent This attraction could stem, for example, from a van der Waals interaction
arising from a non perfect matching between the refraction index of
the solvent and of the polymer in a way that does not alter the basic
configuration of the single star polymer; in this case $w(r)$ is
independent from the temperature.  Alternatively attraction could be
induced by depletion interaction when a third component, which is
large compared with the solvent molecules but small compared with the
star polymers, is present in the solution.  In this case $w(r)$ has an
entropic origin and consequently it depends on $T$. In this situation
the size and concentration of the depletant are, respectively, closely
tied to the range and strength of the attraction.  In what follows we
shall use temperature as control parameter.

 To study the structural arrest and the dynamics
in presence of attractive forces
 we can use the simplified
model potential
utilized in ref.\cite{artmio} where $w(r)$
has the  functional form of a Fermi distribution, i.e.

\begin{equation}\label{Vi}
w(r)=-
\frac{C}{\exp\left[\frac{r-A}{B}\right]+1}.
\end{equation}

\noindent
The parameters $A$ and $B$ control the position and the width of the
well potential, $C$ is the amplitude of the attractive contribution.
For convenience we use reduced units for temperature in terms of $C$: $T^*=k_BT/C$.
By a suitable choice of these parameters one can guarantee,
for the temperatures of interest, that $V_{tot}(r)$ does not have a significant subsidiary maximum at large $r$ and  avoid possible complications due to the competition over different length scales.
We have studied three sets of parameters $A$ and $B$ in order to see the dependence of the glass transition on the range and the width of the attractive well.
The first case corresponds to the
potential discussed in Ref.~\cite{artmio}, i.e. $A=2.1\sigma$,
$B=0.35\sigma$ and we refer to this case as $SP_1$. For
these values of the parameters, the attractive contribution is rather long
ranged and we found that $w(r)$
does not modify the ideal glass transition with respect to
the purely repulsive case.
For the other two choices of the parameter,
$SP_2$ corresponding to $A=1.95\sigma$ and $B=0.21\sigma$ and
$SP_3$ corresponding to $A=1.875\sigma$ and $B=0.155\sigma$ the attractive well is displaced to smaller distance and
we observed significant differences for the glass transition with respect to the only repulsive
pair interaction.
In Fig.\ \ref{fig1} we
show the shape of the potential in the three cases for $f=50$.
It should be noticed that due to the ultrasoft character of $V_{rep}$
the value of
$A$ and $B$ modifies
not only the range of the interaction and the position of the minimum but also the depth of the well potential.
In addition, the position and width of the attractive well  depends on temperature because $V_{rep}$ scales with $T$.
In Fig.\ \ref{fig1} we plot also $V_{rep}$:
at short distances we can observe that $V_{tot}$ is softer than $V_{rep}$.

\subsection{Modified hypernetted chain integral equation}
\label{MHNC-sub}

In the present work we carry out a Mode-Coupling Theory (MCT) analysis \cite{mct} of the
long-time limit of the correlation functions to locate the ideal glass transition line of
the system.  MCT provides a set of closed equations to calculate the non-ergodicity
parameter $f_q$ which acts as an order parameter for the glass (see \ref{MCT-sub}). All
the information needed to solve these equations is contained in the static structure
factor, defined as $S(q)={\langle \rho(-q)\rho(q)\rangle}/N$ and in the number density,
$\rho=N/V$, $\rho(q)$ being the density fluctuation variable of wave vector $q$. To
calculate $S(q)$, we utilized the MHNC integral equation.  This equation is in general
accurate also when an attractive contribution to the interaction is present \cite{MHNC}.
In the case of star polymers in a good solvent it has been verified the remarkable
accuracy of this theory, to describe fluid states as well as metastable states
\cite{artvetri,artmio} for a large range of $f$ and density values.

The starting point for the MHNC equation is an an exact relation
\cite{Hans}, obtained from a cluster expansion, which connects the
radial distribution function (rdf) $g(r)$ to the interparticle
potential $V(r)$:

\begin{equation}\label{ponte}
g(r)={\exp}[-\beta V(r)+h(r)-c(r)+E(r)]
\end{equation}

\noindent
where $h(r)=g(r)-1$ ~and~
$c(r)$  ~are the pair and the direct
correlation function, respectively.
$c(r)$ is related to $h(r)$ by
the Ornstein-Zernike equation.

\noindent
The term $E(r)$, called {bridge function}, represents a sum of an
infinite number of terms, the so called {elementary} graphs in the
diagrammatic analysis of the two-point function and, in general, it is
not known.  In the MHNC scheme the $E(r)$
is replaced by the bridge function of a fluid of hard spheres,
$E_{HS}(r)$, of suitable diameter $d$.  To optimize this choice, which
depends on the parameter $d$, the free energy is minimized
\cite{lado1}.
This is equivalent to satisfy the relation:

\begin{equation}\label{Lado}
\int d{\bf{r}}\left[g(r)-g_{HS}(r,{\eta}_{HS})\right]
\frac{\partial E_{HS}(r,{\eta}_{HS})}{\partial{\eta}_{HS}}= 0
\end{equation}

\noindent
where $\eta_{HS}=\frac{\pi}{6}\rho {d}^3$

\noindent
In order to implement MHNC one needs the rdf $g_{HS}(r)$ of hard spheres from which one can obtain $E_{HS}$.
Verlet and Weis (VW) \cite{A26}
provided an accurate parametrization of $g_{HS}(r)$ based on the PY
equation with a correction which incorporates thermodynamical
consistency through the Carnahan-Starling state equation
\cite{Hans}.
This together with equation (\ref{ponte}-\ref{Lado}) gives a closed set
of equations which are solved
by a standard iterative method.

In a previous paper some of the authors showed that the dependence of $\eta_{HS}$ on the
density as determined by Eq.~(\ref{Lado}) is unusual and reflects the features of the
interparticles interaction \cite{artmio}.  Moreover in ref.\cite{artvetri} we found
evidence that the characteristic sequence of maxima and minima of $\eta_{HS}$ versus
$\eta$ is directly related to the non-monotonic behavior of the diffusion coefficient as a
function of the packing fraction.  Indeed the slow dynamics in star polymer systems at low
and intermediate densities can be qualitatively described as the slow dynamics of the hard
sphere system via a density and functionality dependent effective diameter determined with
MHNC closure.

\subsection{Molecular Dynamic simulation: comparison between MHNC and MD
results on the correlations.}
\label{MD-sub}

To check the results of the MHNC and MCT calculations, we performed extensive MD
simulations for the model described by the potential defined by Eq.~(\ref{Vot2}-\ref{Vi}).
We simulate $N=1000$ particles for different values of the density, temperature and
functionality. For each state point the configurations have been equilibrated at constant
temperature for a time long enough to ensure both the equilibration and decorrelation from
the initial configuration. The acquisition run started after this preliminary preparation,
then the system was simulated at constant energy. We carefully checked whether
crystallization occurred or not during the run by inspection of the static structure factor,
with the same modulus, not averaged over different directions of $q$-vectors. In fact to
improve the averages in the calculations of $S(q)$ at a given $q$, we generally considered
up to $300$ independent $q$-vectors chosen with a random direction but with the same
modulus $q$. In correspondence of these values, we evaluated the density variables and
consequently the static structure factor for a given direction. Finally we averaged on the
different directions. For a liquid the structure is extremely disordered and all the terms
for different directions give roughly the same contribution to the average. When the
structure factor corresponding to a certain direction starts to grow over the others, the
system begins to show a preferable direction, i.e. crystallization  takes place. When a
configuration crystallized we discarded it. As discussed for the purely repulsive case
\cite{artvetri}, the system has a strong tendency to crystallize, being monodisperse in
diameter \cite{foffi}. Consequently, we will test our predictions only when the system
remains in the liquid phase, a future direction of our research may be directed toward the
suppression of crystallization introducing, for example, a second component slightly
different in diameter \cite{foffi}. We chose as unit of length the corona diameter
$\sigma$ and as unit of mass the mass of the particles. Moreover we measure the
temperature in reduced units, i.e. $T^*=k_BT/C$.

We begin our discussion on the numerical simulation by discussing the
purely repulsive case, i.e. $C=0$. As extensively described in
ref. \cite{artmio}, MHNC gives a very accurate description of
correlations for star polymers in a good solvent for a wide range of
densities. In Fig.\ \ref{fig2} we present results for the repulsive
case, for $f=32$ and $\eta=0.6$ obtained from MHNC and Rogers-Young
(RY) theoretical calculations and from MC
and MD simulations
\cite{57footnote}
The RY equation is another integral equation for $g(r)$ which
interpolates between PY and HNC equations \cite{RYY}.  The agreement
between MD and MHNC, and between MD simulations and MC simulations, is
extremely good.  The small apparent MD underestimation of the main
peak is due to the grid we used in the simulations. A more refine
grid, however, would increase the noise in the data.  On the other
hand RY shows some significant discrepancies with respect to
simulations results in the range of strong coupling which emerges for
packing fractions of order of $\eta=0.60$ and $\eta=3$
\cite{artmio}. By comparison with simulation results (both MC and MD)
we definitively conclude that MHNC describe accurately the structure
of star polymers in a good solvent.  

Hence, we focused on the attractive case and we compare the MD results
and the prediction of the MHNC theory for the fluid case:
in particular we compared  the radial distribution function and the
structure factor for the three different models of the attraction
described in section \ref{MHNC-sub} in a regime of strong coupling
when the main maximum of $S(q)$ reaches large values.
The temperatures and packing fractions
we have
chosen in our study, correspond to states of interest in the study of the ideal
glass transition line, i.e.
close to the fluid-glass and glass-fluid ideal transition line.  We
find in general a very good agreement between MD results and MHNC
results for all the values of temperature, packing fraction and arm
number investigated. In particular a good accuracy is achieved
in the determination of the position of the main peak in the radial
distribution functions and structure factors. Only for states at low
temperature and at packing fraction corresponding to
very large coupling
(e.g. see Fig.\ \ref{fig4}) one notices some small discrepancies
between MHNC and MD results.

This analysis confirms
the
accuracy of the
MHNC integral equation in describing the structure of an
ultrasoftcore potential also in presence of an attractive
contribution at longer range.

One can notice
that the MHNC $g(r)$ has some structure in the
region of the second maximum and this can even show up as a subsidiary
maximum (Fig.\ \ref{fig4}). It is known \cite{MHNC} that this
spurious structure is a consequence of the VW parameterization of
$g_{HS}$ in terms of the PY solution but it is believed that this
anomaly has no serious consequence in $S(q)$.

\subsection{Mode Coupling Theory for the Ideal glass transition}
\label{MCT-sub}
In this section, we shall briefly review the nature of MCT, and
discuss the type of information it yields. The MCT of super-cooled
liquids describes the dynamical transition by a nonlinear
integro-differential system of equations for the normalized time
correlation functions of density fluctuations $\Phi(q,t)= {\langle
\rho(-q,t)\rho(q,0)\rangle}/ {\langle \rho(-q,0)\rho(q,0)\rangle} $,
where $\rho({\bf q},t)=\sum_l \exp{(i {\bf q}\cdot {\bf r}_l(t))}$. As
discussed above, the only input to the MCT equations are the
equilibrium static structure factor, $S(q)$ and the number density,
$\rho$. The glass transition  can be identified by studying the
long time limit of the MCT equations, which determine the
non-ergodicity parameter of the system $f_q=\lim_{t\rightarrow
\infty}\Phi(q,t)$. An ergodic state is characterized by $f_q=0$. This
value is always a solution of the MCT long-time limit equations
\cite{mct}.

The quantity $f_q$ obeys the equation $f_q/(1-f_q) = {\mathcal
F}_q(f)$.  Here, the mode-coupling functional ${\mathcal F}_q$ is
given by\\
\label{eq:functional}
\begin{equation}\label{eq:functional-a}
{\mathcal F}_q(f)  = \frac{1}{2} \int{ {{{d}^3k} \over
(2 \pi )^3} V_{\vec{q},\vec{k}} f_k f_{|\vec{q}-\vec{k}|}}\,.
\end{equation}
\\
Equation \ref{eq:functional-a} together with the equation for $f_q$ can be
derived by taking the long time limit, $t\rightarrow \infty$, of the
MCT equations \cite{mct}.  The mode-coupling vertices are determined by the
structure factor $S_q$, the direct correlation function $c_q$, and the
density $\rho$:
\begin{equation}\label{eq:functional-b}
V_{\vec{q},\vec{k}} \equiv  S_q S_k S_{|\vec{q}-\vec{k}|} \rho
\left[ {\vec{q}} \cdot
\vec{k}\,{c_k} +\vec{q} \cdot
(\vec{q}-\vec{k})\,{{c_{|\vec{q}-\vec{k}|}} }
 \right]^2/q^4
\;.
\end{equation}
The direct correlation function is directly related to the static
structure factor by the relation
$c_q=\frac{1}{\rho}\left(1-\frac{1}{S_q}\right)$. The glass
transition appears as an ergodic to non-ergodic transition for the
system, where $f_q \neq 0$ solutions arise. These transition points
correspond to bifurcation singularities of the MCT
Eq.~(\ref{eq:functional-a}-\ref{eq:functional-b}).\\ In the present
work, we numerically solved
Eq.~(\ref{eq:functional-a}-\ref{eq:functional-b}) with an iterative
procedure over a grid of $650$ equi-spaced q-vectors up to
$q=62.32$.  For the static structure factor we used the static MHNC
structure factor $S(q)$ calculated as described above.

\section{Results}
\label{RESULTS-sec}
\subsection{$SP_1$: MCT and MD results}
\label{att1-sub}


First of all we considered the model potential we labeled $SP_1$. At low temperature the
system has fluid-fluid phase transition with two critical points: the first critical point
is around $\eta_c=0.026$ and $T^{*}_c\simeq0.7$, the second one around $\eta_c=1.04$ and
$T^{*}_c\simeq0.156$. Far from the critical point, we observed that the structure factor
$S(q)$, relative to the potential $V_{tot}$, is not very different from the structure
factor relative to the simply repulsive interaction $V_{rep}$. In particular the position
of the main peak does not change appreciably and its height change for less than $1\%$
($0.2 \lesssim T^{*}\lesssim0.6$). Consequently one might expect that the location of the
ideal glass transition will not be different from the repulsive case~\cite{artvetri}. Indeed this is the
case.   We studied several arm numbers (from low value, $f=24,32$ to higher one $f=70$)
for several values of the packing fraction with MCT.  Similarly to the case of purely
repulsive interaction we  deduce that below $f=46$ there is no glass phase. 
For $46<f<60$ the system is 
fluid for low densities and glass for intermediate ($0.25 \lesssim
\eta\lesssim 0.7$), but it is still fluid at high densities
($\eta>0.7$).
When $60<f<72$ the system is in a glass state for intermediate and
very high densities ($0.25 \lesssim \eta\lesssim 0.7$, $\eta\gtrsim
2.25$).  Finally above $f\simeq72$ the system is fluid for low
densities ($\eta \lesssim 0.25$) and in a glass state for intermediate
and high densities ($\eta\gtrsim 0.25$).

Hence, the effect of the attraction does not change the location of
the ideal glass transition. Consequently, in the temperature-packing
fraction plane, the ideal glass transition line would be trivially
represented by a vertical line.

We also performed molecular dynamic simulations to calculate the
dependence of the diffusion coefficient, from the long time limit of
the mean squared displacement, varying the temperature, for different
values of $f$ and the packing fraction.  This analysis supported the MCT
predictions, i.e. the diffusion coefficient  does not change
appreciably with the temperature.


Starting from the behavior of the structure factor changing the
temperature and from the analysis of the structural arrest we
performed, we can complete the picture of the phase diagram. For
$f\leq 34$ there is no freezing transition: the two fluid-fluid phase
transitions represent stable states; for $f\gtrsim 50$ all the density
region above the density corresponding to the triple point $T_p$ is
occupied by crystalline phases; for $34<f\lesssim50$ bot the
fluid-fluid phase transitions and their critical points 
 persist as stable states. In Fig.\ \ref{fig6} we describe the
whole phase diagram for $f=50$: the fluid-fluid phase diagram is
calculated by means of mean field theory and the ideal glass
transition lines determined with MCT.  At the equilibrium the region
around the second critical point becomes metastable with respect to
the freezing, the regions between two successive squares (from low to
high densities), in the upper part of the figure, indicate the
densities where the system is solid (data are MC results reproduced by
Watzlawek {\it{et al.}} \cite{Phtd}). If the crystallization can be
avoided we can observe that the second critical point lies outside the
glassy region (see also the inset).  The limit of very high
temperatures, i.e. the purely repulsive interaction are also
shown in Fig.~\ref{fig6}: the densities which delimit the glass
region are effectively the same with and without attractive
interaction.

We conclude that, varying $f$, the first critical point will always lie 
outside the glass region. The second critical point, however, will
enter the glassy phase for $f\gtrsim 70$.

\subsection{$SP_2$: MCT and MD results}
As anticipated above we tried to tune the parameters A and B in order to
enhance the effect of the attraction on the shape of the
glass-transition line.  Our aim was to shrink the well and move the
minimum of the pair interaction close to sigma.

First of all we consider the parameters $A=1.95\sigma$ an $B=0.21\sigma$
($SP_2$).  As we  see in Fig.\ \ref{fig1} the well potential in
this case  is shrunken  approximately by $30\%$
with respect to $SP_1$
and the
position of the minimum is changed approximatively by $6\%$
($0.25<T^*<0.6$). If compared to $SP_1$ the depth of the well potential changed from
$5\%$ with respect to at lower temperature till
approximatively $28\%$ around
$T^{*}=0.6$.

\noindent
In Fig.\ \ref{fig7} we present the occurrence of the glassy phase for
different numbers of arm and as a function of temperature.  For
$f>46$ we can summarize our results as follows: for high temperatures
the two ideal-glass lines (fluid-glass on the left and glass-fluid on
the right) tend to the repulsive case~\cite{artvetri}.
On lowering the temperature, however, a fluid-stabilizing effect sets
in, so that the fluid-glass line tend to move to larger values of the
density.  So in the low density region, roughly below $\eta=0.50$,
there is the possibility of the glass melting when lowering the
temperature.  In this case increasing the temperature, the width of
the nonergodicity parameter $f_q$, which is a measure of the inverse
of the cage localization length, gets larger (see Fig.\ \ref{fig8}).
Moreover we observe an increase of $f(q=0)$ lowering the temperature,
corresponding to densities closer the coexistence curve. The effect of
the attraction for large value of $f$ is very small and in particular
for $f=70$ we can see that the fluid-glass line is again very similar
to the repulsive case.  On the right side of the glass region we can
observe that the glass-fluid line move to higher densities when the
temperature decreases and consequently when the intensity of the
attraction increases.  In contrast to the low density case, the
attraction now favors the formation of the glass to higher density
with respect to the repulsive case.  On lowering the temperature the
nonergodicity parameter $f_q$ presents a larger width in $q$. This
clearly indicates that the particles are localized on a shorter length
scale.  This phenomenon could be related either to the effect of a
stronger attraction or to a larger value of the effective repulsive
length. It is interesting to stress that in all the cases taken into
account the non-ergodicity parameter, presents the typical
shape of a repulsive glass, i.e. a glass that possesses a structure
dominated by the caging effect. This typical repulsive behavior is
characterized by oscillations in correspondence to the peaks of the
static structure factor with a maximum at the first peak. It has been
shown, both by MCT calculations \cite{Dawson2001} and computer
simulations
\cite{Zaccarelli2002b}, that for sufficiently short-ranged attraction
 a new glass, that has been named attractive glass, emerges. Indeed
 the shape of the $f_q$ for an attractive glass is completely
 different: the oscillations are very weak and the maximum is not so
 pronounced. In our investigation we encountered fingerprints only of
 the repulsive glass. The effect of the attraction and the interplay
 between attraction and repulsion, seem to act more on the size of
 the cages and on their formation rather than on the nature of the
 arrest itself. However we have not investigated the case of very
 narrow attractive wells on the length scale of $\sigma$.

The behavior of star polymers with $f=45$ is different: we can
observe that the attractive contribution of the interaction favors
the occurrence of glass states at low temperature.  Indeed for this
value of the functionality, a stable glass phase emerges at a
temperature $T^{*}\lesssim0.6$ from the solution of MCT equations and
the density range of the glass phase gets larger on further decrease
of the temperature.
In this case the q-width of $f(q)$ increases lowering the temperature, both for  $\eta<0.5$ and $\eta>0.5$.
 It is interesting to note that no glass transition
is found in the corresponding repulsive case, i.e. in the limit of
high temperature.

In the so called ``repulsive glasses" the occurrence of the ideal
glass transition in the framework of the MCT  depends  strongly on
the behavior of the main peak of the structure factor, i.e. on the first neighbor interactions.

\noindent
In Fig.\ \ref{fig10bis},
we can observe a magnification of the main peak
of the structure factor for $f=50$ and two different packing fractions
on the left and on the right side of the ideal glass region (MD
simulations).  For the different temperature investigated we observe
the same trend: for $\eta=0.314$ (Fig.\ \ref{fig10bis}-$a)$)  decreasing the temperature the main
peak of the structure factor decreases, showing a loss of the
correlation between particles.  The opposite trend is observed on the
right side, $\eta=0.628$ (Fig.\ \ref{fig10bis}-$b)$).  We also studied 
the behavior 
of the first
peak in the structure factor for $f=45$, in this case the trend in the
glass region is represented by an increase of the main peak lowering
the temperature, indeed this effect is responsible for the anomalous
formation of the glass at this value of the functionality.

This preliminary and qualitative study, has been supplemented by a
more thorough analysis: we performed molecular dynamic simulation for
several values of $f$, $T^*$ and $\eta$.

\noindent
In Fig.\ \ref{fig9} we show the diffusion coefficient $D/D_0$ as a
function of the temperature calculated for packing fraction not far
from the glass region and for different values of $f$
($42,45,46,50$). In the present context data for $D$ are normalized by
$D_0=\sigma \sqrt{T^*/m}$, in order to take into account the
temperature dependence of the microscopic time~\cite{foffi}.  The
MD-diffusion coefficients as a function of $T^*$ confirm the MCT
trend.  On the low density side of the glass region the diffusion
coefficient decreases increasing the temperature while on the high
density side the diffusion coefficient increases increasing the
temperature.  
The values of $D/D_0$, for fixed temperature,
 decrease increasing
$f$; we remark that
for $f=42$
the system does not show
glass transition.

\noindent In general it has been observed that the effect of the presence of a glass
transition can be noticed as a decrease in the diffusivity also far
from that part of the phase diagram where the structural relaxation
time starts to grow. Indeed this has been encountered, for example, in
a monodisperse square well system~\cite{foffi} and in a purely repulsive
soft potential~\cite{artvetri}. In MD simulations both these systems
presented a strong tendency to crystallize.
However, studying the
behavior of the diffusivity, it is possible to pin down the shape of
the glass transition line. Also in the present work we managed to
check that the diffusion follows the MCT predictions. The fact that
the $f=45$ trend could not be checked, i.e.  an increase of the
diffusivity on rising the temperature for both high and low density
region, could be related to the fact that the simulations should be
run closer to the glass transition and, unfortunately, this is not
feasible due to the occurrence of the crystallization.  One way around
this problem would be to extend the results to a system that presents
a smaller crystal nucleation rate. For the square well system, for
example, this has been accomplished considering a binary
mixture~\cite{Zaccarelli2002b}.

Summarizing we remark, for the different arm numbers investigated,
that for $\eta\lesssim 0.5$ the attraction seems to determine a
destabilization of the cages while for $\eta \gtrsim 0.5$ the attraction
facilitate the formation of the glass for higher density with respect
to $V_{rep}$.  The most peculiar behavior has been found for a value
of $f$ ($f=45$) for which the glass lines are very near to the value
of $\eta=0.5$.  For this packing fraction star polymers start to
interpenetrate widely.  We emphasize that, considering stars in a good
solvent at the equilibrium, this value is just in the middle of the
bcc ($34<f\lesssim55$) or fcc ($f>55$) crystal phase.  Related to the
behavior of the structure factor corresponding to $SP_2$, we expect
that also the densities where the system is solid, at the equilibrium,
could change weakly with respect to the repulsive case.  We will
return more in detail on this point in the section relative to
conclusions and discussion.

Finally we show in Fig.\ \ref{fig13} the position of the ideal-glass
lines with respect to the fluid-fluid coexistence curves for $f=50$.
It turns out that the temperature should be decreased a lot to notice
possible effects of the density fluctuations around the second critical
point on the ideal glass transition line.

\subsection{$SP_3$: MCT and MD results}
The third case
we considered is
characterized by $A=1.875\sigma$ and
$B=0.155\sigma$, and we indicated it $SP_3$.
In this case the width of the well potential decreased, with respect to
$SP_1$,
from approximatively   $40\%$ at $T^*=0.25$ till $60\%$
at $T^*=0.6$.
The position of the minimum
is changed by less than $9\%$.
If compared to $SP_1$ the depth of the well potential changed
from
$7\%$ with respect to at lower temperature till
approximatively $48\%$ around
$T^{*}=0.6$.

In Fig.\ \ref{fig14}  the occurrence of the glass phase is shown
for $f=46,50,70$  for a wide
range of temperatures.
We analyzed in more detail the case $f\leq70$ as here we expect very different
behaviors from the three cases in exam.

For $f\geq50$ the qualitative behavior is very similar to that of
$SP_2$. For densities below the glass region the influence of the
attractive term, when lowering the temperature, is a
liquid-stabilizing effect.  In the opposite regime (densities above
the glass region), increasing the intensity of the attractive
contribution results in moving the glass-fluid line to higher
densities with respect to $V_{rep}$.  We point out that here the
curvature of the lines is more pronounced than in the $SP_2$ case.
Anyway comparing all the cases investigated is not trivial: this is in
part due to the fact that a correct rescaling of the temperature
should be done considering the change in the depth and the position of
the well potential changing $A$, $B$ and $T$, rather than the
amplitude $C$ of the attractive term. We will return on this point in
section\ \ref{Concl-sec}.

\noindent
For $f\leq 46$
the behavior of the star polymer solution is completely different from the previous cases.
Indeed for $f=46$ it exists a value of the temperature, $T^{*}\simeq 0.6$,
below which the system
does not present a glass transition.
For  $f=45$ we do not observe any glass phase for all the temperature values
investigated ($0.4\gtrsim T^*\gtrsim 0.8$).

As for the previous case, we performed MD simulations for different
arm numbers and temperatures, confirming the MCT result: the trend of
$D/D_0$ is qualitatively the same as for $SP_2$ (see Fig.\
\ref{fig15}).  For $f=46$ we do not capture the increasing of the
diffusivity when decreasing the temperature, as we expect on the basis
of the MCT results on the right side of the glass region.  As in the
previous case this might be due to the fact that we did not consider
packing fraction close enough to the ideal glass lines.

Finally we studied the behavior of the structure factor, changing the temperature,
for
$f\leq 50$.
As
for $SP_2$, we calculated
$S(q)$
on the left and on the right side of the ideal glass region (MD
simulations).

\noindent
For $f=50$ the trend is the same observed for $SP_2$, i.e.  on
decreasing the temperature for $\eta=0.314$  the main peak of the
structure factor decreases, showing a loss of the correlation between
particles.  The opposite trend is observed on the right side
($\eta=0.628$).

\noindent
For $f=46$ the behavior of the structure factor is different.  In
Fig.\ \ref{fig16}-$a$, we present a magnification of the main peak in
the structure factor for $f=46$ and $\eta=0.55$ (right side of the
glass region).  The picture shows a decrease of the main peak of
$S(q)$ lowering the temperature.  Moreover in Fig.\ \ref{fig16}-$b$ we
present a magnification of the main peak in the structure factor for
the same arm number and $\eta=0.51$, (left side of the glass region).
The trend is exactly the same as for $\eta=0.55$: we stress that,
introducing attractive forces, star polymer solutions of $46$ arms,
show a loss of correlations among particles which are responsible for
the peculiar MCT predictions (Fig.\ \ref{fig14}).  This effect is
enhanced as the interaction is increased.

\noindent
Moreover we also  studied 
the behavior of the structure
factor for $f=45$ and $\eta=0.51$, in this case the trend is
represented by a decrease of the main peak lowering the temperature.
We  observed that the height of the main peak introducing the
attractive contribution and increasing the intensity of the attraction
is in any case smaller than the peak height relative to $V_{rep}$.  We
recall that for $f=45$, MCT does not give a glass phase neither for
$V_{rep}$ nor for $SP_3$.  

Summarizing we can argue from Fig.\ \ref{fig14} that for the model
interaction $SP_3$ when $\eta \lesssim 0.5$ the attraction seems
always to determine a destabilization of the cages, while for packing
fractions larger than $0.5$ the attraction facilitates the formation
of the glass for higher density with respect to $V_{rep}$.  For $\eta
\simeq 0.5$ (corresponding to the
region of interest for $f=46$)
the effect of the attraction seems to determine a
destabilization of the cages only.

In Fig.\ \ref{fig18} we present the fluid-fluid phase diagram and the
glass transition lines for $f=50$.

We conclude this section remarking that, so far, we discuss about the
behavior of the first peak in $S(q)$ only, while the second peak in
the structure factor does not change appreciably in all the cases
investigated.

\section{Conclusions and Discussion}
\label{Concl-sec}

In this work, we studied the structural arrest and the dynamics in
star polymer solutions when attractive forces between macroparticles
are present.  The model potential we used to describe the interactions
presents an ultrasoft repulsive term of entropic origin at short range
plus an attractive interaction at longer range.  Due to the addition
of the attractive contribution between star polymers the repulsive
core becomes softer on lowering the temperature and consequently
increasing the intensity of the attraction.  We analyzed three
different forms of the pair interaction between stars, considering
attractive forces of shorter and shorter range.  In this paper we
focused on stars with several arm numbers (from low value, $f=24,32$
to higher one $f=70$) at several values of density and temperature.

We examined the structure of the solution solving MHNC closure and
performing extensive MD simulations, in this way we tested the
accuracy of the MHNC to describe the properties which arise from
such choice of the interaction model.  Over the full range of
densities and temperatures of interest MHNC and MD simulations are in
a good agreement. Hence we have been able to conclude that MHNC is a
good approximation to study systems described by ultrasoft-core
repulsive interactions, with and without attractive forces between
macroparticles.
Having tested the accuracy of the MHNC approach, we focused our
attention on the location of the ideal glass transition, studied
within MCT.

In particular for the case characterized by the longer range of the
attraction, named $SP_1$, the ideal-glass transition line we obtained
is not significantly modified in comparison with the one obtained for
the purely repulsive potential.  Indeed both the structure and the
dynamics of the solution are not significantly modified by introducing
the attractive contribution. On the other hand the two other systems
characterized by a shorter attractive range show some significant
differences with respect to the purely repulsive pair interaction.

For $f\gtrsim50$ $SP_2$ and $SP_3$, show
the same qualitative behavior.  In particular we can distinguish two
regions in density, a low density regime ($\eta<0.5$) and a high
density regime ($\eta>0.5$) where the system, in both cases, behaves
differently. For low densities, on lowering the temperature, a
liquid-stabilizing effect due to the attractive forces sets in, so
that the liquid-glass transition line moves to larger values of the
density.  Indeed, this is an interesting effect since it presents the
possibility to pass from a glass phase to a liquid phase decreasing
the temperature. 
It is perhaps worth making a few remarks about this
issue. It is now well established that systems characterized by a step
repulsion and a short range attraction, possess a reentrance in the
the dynamical arrest. In these systems it is possible to melt the
glass by lowering the temperature. This is a phenomenon now well
established in theory, simulations and experiments
\cite{P1,P2,P3,P4,ken}. Clearly, one would be tempted to relate our
findings to this phenomenology of colloidal solutions; however we
showed that the origin of such effect in our case is different. For
short range attractive colloidal systems at low temperature there is a
glass phase originated solely by attraction and the reentrant melting
arises from the competition between the high temperature and low
temperature regime.  For our system, however, there is no indication
of an attraction dominated glass at low temperature and the
destabilization of the glass is due to the destabilization of the
cages that, in the high temperature regime, are responsible of the
arrest. For $\eta\gtrsim0.5$, we found that the glass-fluid line moves
to higher densities when the temperature decreases.  As consequence
increasing the intensity of the attraction, for a fixed density and
arm number, the system moves from a liquid to a glass.  This behavior
is closely connected to the ultrasoft nature of the pair
interaction. We recall that stars in a good solvent shows a remelting
of the glass phases (or solid phases at the equilibrium) for very high
densities. The addiction of attractive forces moves the glass-fluid
line to higher densities. We could explain this effect as a
contribution of the attraction to stabilize the cages and then to
inhibit the remelting of the solution.

For $f=45,46$ the behaviors of $SP_2$ and $SP_3$
are very different. In the case $f=46$
for $SP_2$  we  observed
the  qualitative behavior found for $f\geq50$,  while
for $SP_3$ there exists a value of the temperature, $T^{*}\simeq 0.6$,
below which the system
does not present any sign of a glass transition.
For  $f=45$ considering $SP_2$
we observed the presence of a glass transition,
 while for the simply repulsive interaction
this transition is not present for such value of $f$, i.e.
the attraction seems to facilitate the formation of the glass.
In the $SP_3$ case we did not observe glass phases for all the temperature values
investigated.
So in this case the effect of the   addition of the
attraction
is a destabilization of the cages only.

To verify the MCT results  we performed the analysis of the dynamics by means
of MD simulations.
As introduced  in section\ \ref{att1-sub}, the range of density
which we examined
is not
sufficiently close to the ideal
glass lines.
For density closer to the transition lines
(mainly for  $f=45,46$)
the system show a strong tendency to the crystallization.
So we verify our MCT predictions for $f\geq 50$.

\begin{figure}[ht]
\centerline{\psfig{figure=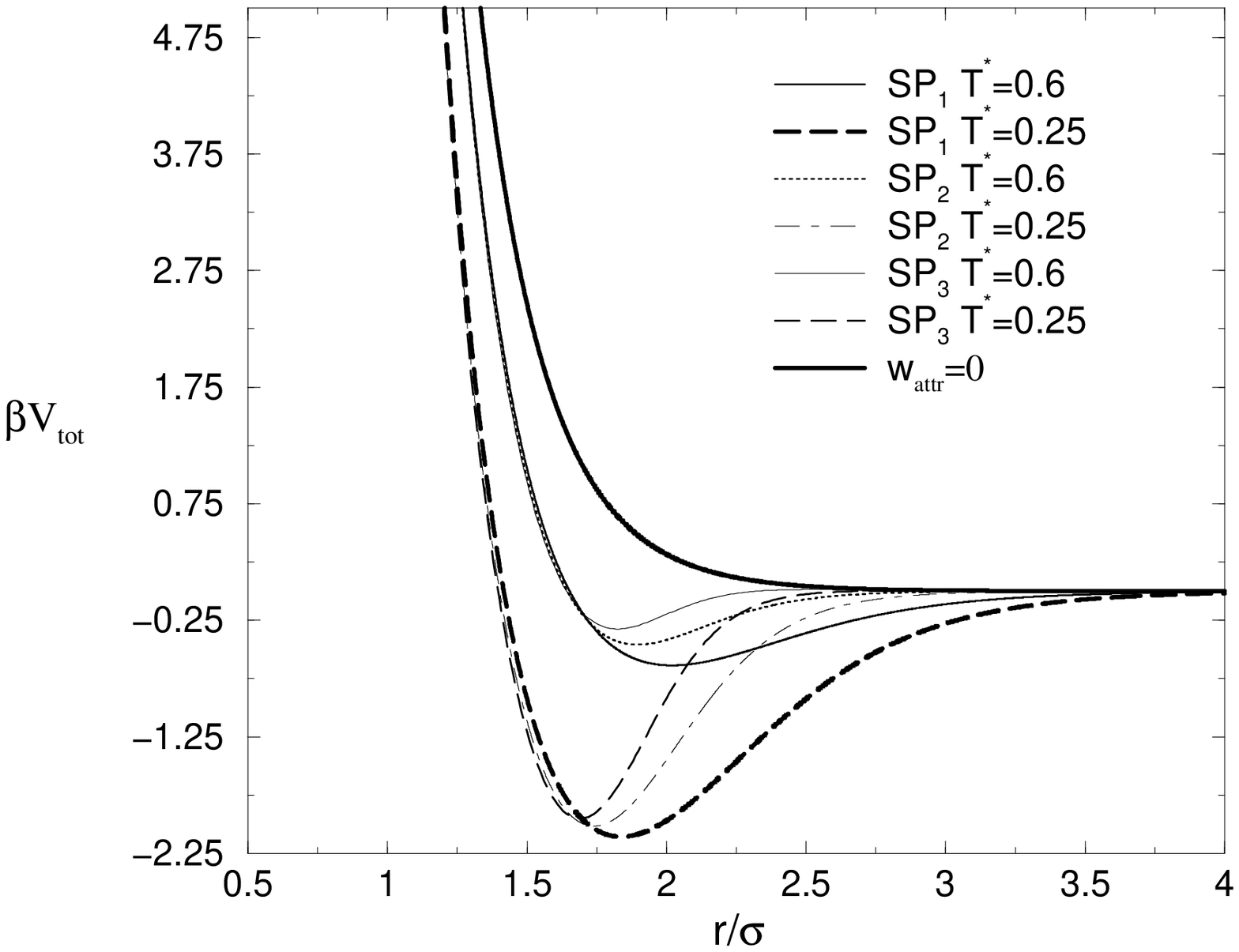,height=10cm}}
\caption{}

\label{fig1}
\end{figure}

\begin{figure}[ht]
\centerline{\psfig{figure=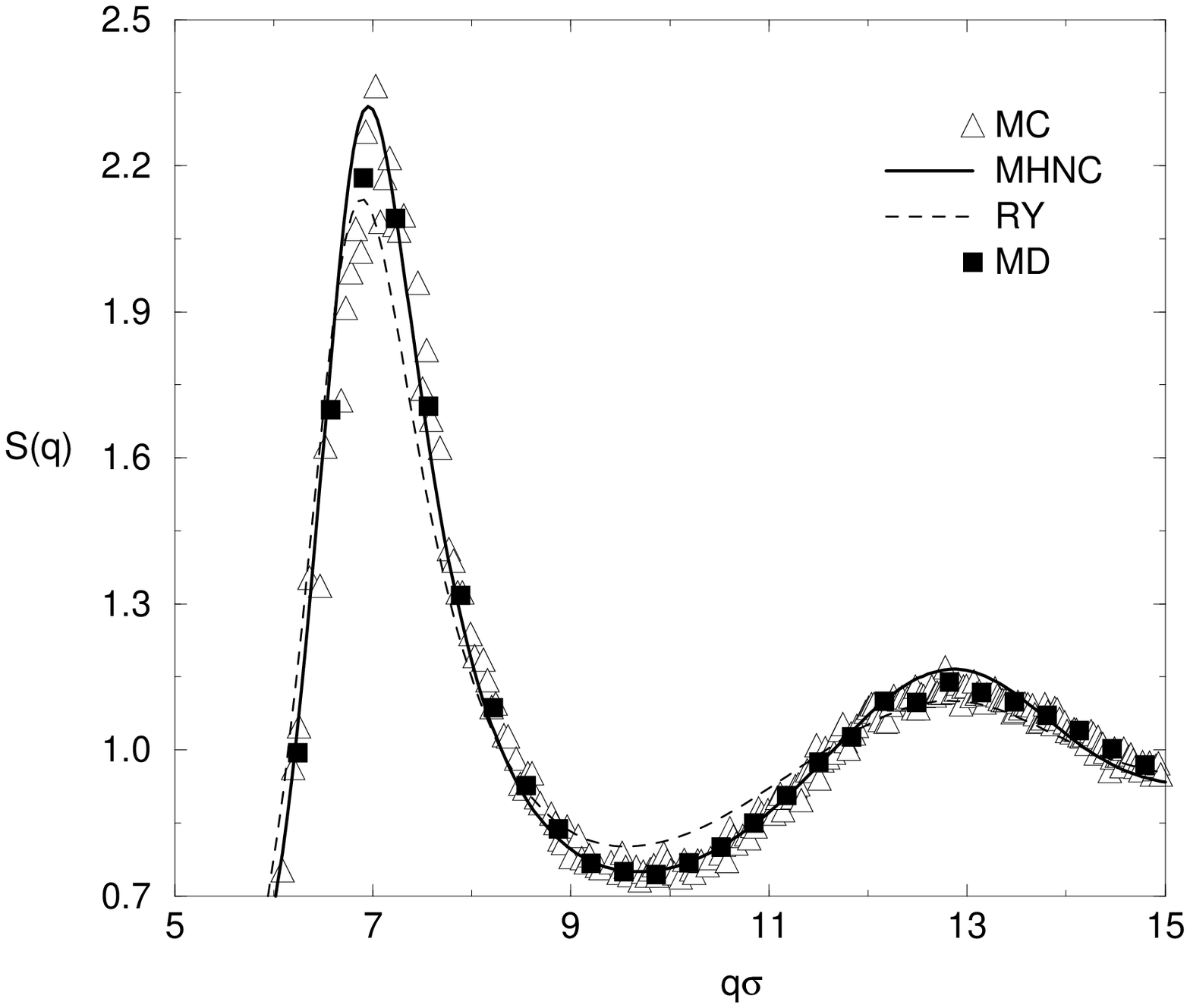,height=10cm}}
\caption{}

\label{fig2}
\end{figure}

\begin{figure}[ht]
\centerline{\psfig{figure=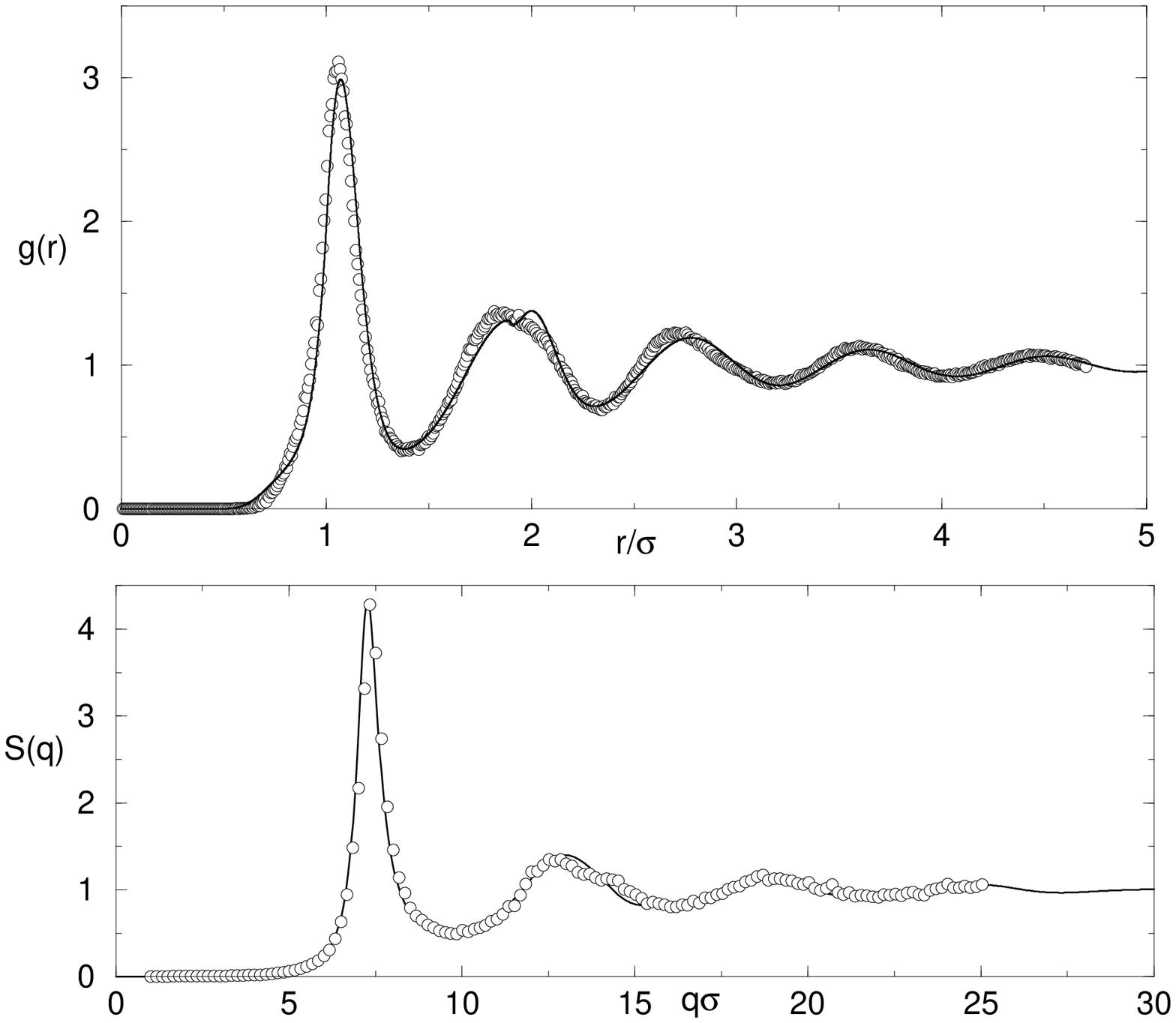,height=12cm}}
\caption{}

\label{fig4}
\end{figure}

From the whole analysis performed, we evidenced a value of the packing
fraction, i.e. $\eta=0.5$, which marks a change in the behavior of the
solution.  This value could be traced back to the cross-over between
the two different functional forms of the repulsion: the logarithmic form
and the Yukawa one.  For star polymers in a good solvent $\eta=0.5$,
the so called {\it overlap packing fraction}, corresponds to the
packing fraction above which the radial distribution function show a
coordination shell inside the logarithmic core.  In other words for
$\eta>0.5$ stars start to interpenetrate widely.  Also in this case
the cross-over designates approximatively the transition of the system
through two different regimes.  For $\eta < 0.5$ the response of the
system to the introduction of attractive forces is a destabilization
of the cage.  This is mainly due to the change in the repulsive
contribution at short distances which become softer and softer
considering respectively $SP_1$, $SP_2$ and $SP_3$.  For $\eta>0.5$
stars interpenetrate more and more.  The effect of the attraction
seems to inhibit the remelting of the solution and the glass-fluid
line move to higher densities.  The shift of the transition line could
be understood considering the effect of the attraction on the second
shell of neighbors.  Indeed if we look the radial distribution
function very close to the glass-fluid transition line, at low
temperature, we observe that the second shell of particles is around
$r=2$.  In this region, see Fig.\ \ref{fig1} we can observe that the
system feels stronger attractive forces passing from $SP_1$ to $SP_2$
and then to $SP_3$.  We conjecture that this attraction on the second
neighbor shell is at the origin of the extended stabilized region of
the glass phase for higher densities.

We have to remark some details relative to our analysis: first of all
$V_{tot}$ does not present considerable repulsive maxima for large $r$
when the temperature is lower than $T^*\lesssim1.$ for $SP_2$ and
$T^*<0.8$ for $SP_3$.  Obviously the repulsive contribution at longer
range is more pronounced for higher temperature (when the intensity of
the attraction diminishes) and low arm numbers (when the Yukawa pair
interaction decay to zero very slowly).  Anyway, also below these
temperatures, where the pair interaction shows small repulsive
contribution at large $r$ (around $\simeq 3.$), we verify that our MCT
predictions do not depend on the existence of the repulsive shoulder.
Indeed we performed the same MCT calculation with a truncated
potential in which the repulsive shoulder is suppressed.  We conclude
that for the temperature and densities of interest in our study the
presence of this small repulsive contribution at large distance does
not alter the picture of the glass regions.  We emphasize that the
presence of the shoulder is not an effect of our particular choice of
the attractive contribution: due to the ultrasoft-core repulsion and
to the Yukawa repulsive contribution at long range the introduction of
attractive forces of shorter range (i.e. depletion forces) could
determine an additional repulsion outside the core.

Since the potential presents an ultrasoft-core interaction, it is
difficult to determine a natural  scale of energy (and length
scale).  We decided to rescale the temperature with respect to the
integrated intensity of the attractive contribution ($T_{new}$) in
such a way to compare our results in a more significant way.  

\noindent
In Fig.\ \ref{fig19} we present the comparison between the MCT data
obtained for $45\leq f\leq 50$ considering the $SP_2$ and $SP_3$
models.  As we can observe for $f=50$ the effect of destabilization of
the cages as well as the inhibition of the melt is more accentuated
for $SP_3$.

Finally $SP_2$ and $SP_3$ show an apparent conflicting behavior
concerning $f=45,46$.  Due to the very small range of $f$ it is very
difficult to understand this peculiar behavior starting from the pair
interaction.  As we remind above, $\eta=0.5$ is a value for which star
polymers in athermal solvent, at the equilibrium, are in a solid phase
and in particular $0.5$ is just in the middle of the solid region
($34<f<70$).  Around this region we find, that the system has a very
strong tendency to crystallize.  Indeed, despite the change observed
in the structure factor, and considering the connection between
equilibrium phase diagram and glass transition line (\cite{artvetri}),
we expect for $SP_2$ and $SP_3$ a small shift of the solid region with
respect to the purely repulsive case.  Notice that this region
corresponds to the glass region we determined for stars with
$f=45,46$.  It is not surprising that our data in the case of
$f=45,46$ show an ambiguous trend between $SP_2$ and $SP_3$.  For the
above mentioned reasons we do not believe it is very interesting to
analyze further on $f\lesssim 46$.

We conclude from this analysis that the details of the phase diagram,
concerning glass transition, is very sensitive to the particular,
specific form of the attractive contribution.  This could be mainly
connected to the change of the short range repulsive contribution when
attractive forces between stars are introduced and not dependent on
the choice of the specific model.  We expect the same behavior also
considering depletion forces.  Moreover in the case of a general
mixtures of micelles with hydrophobic group at the ends of the polymer
chains, as discussed in section \ref{METHsec}, the attractive
contribution will be at very short range and close to interparticle
separation equal to $\sigma$.  This determines a change in a repulsive
interaction in the region where in star polymer solution there is a
cross over between the logarithmic form and the Yukawa one.  We decide
to complete our analysis considering a specific system described by
specific attractive forces.  Starting from this analysis a future
perspective of our work on star polymer solutions is a more direct
comparison between theory and experiments about the origin and the
description of the attractive interaction.  In this sense we decided
to turn our attention, more in general, on systems of micelles which
can be described by soft-core potential plus attraction at shorter
range, similar to star polymer macromolecules.  There exist many
reasons to further study star polymer solutions: a precise
understanding of their properties will give the possibility to make
progress in the exploration of `hybrid' polymeric-colloidal materials
such as irregular multiarm stars, self-organized structures resulting
from telechelic linear homopolymers and copolymers, polyelectrolyte
brushes, micelles with chemically fixed cores.

\vspace{10 mm}

\begin{center}
{\bf Acknowledgments}
\end{center}

\noindent
Funded in part by a grant of 
the Marie Curie program of the European Union, 
Contract Number : MRTN-CT2003-504712.
F. Lo Verso acknowledges financial support from INFM.
G. Foffi and P. Tartaglia
acknowledge financial support from MIUR-COFIN and
MIUR-FIRB and thank Francesco Sciortino for useful discussions.

\clearpage

\clearpage

\centerline{\bf FIGURE CAPTIONS}
\vspace{4mm}

\newcounter{nuovo}
\baselineskip 7.8 mm

\begin{figure}[ht]
\centerline{\psfig{figure=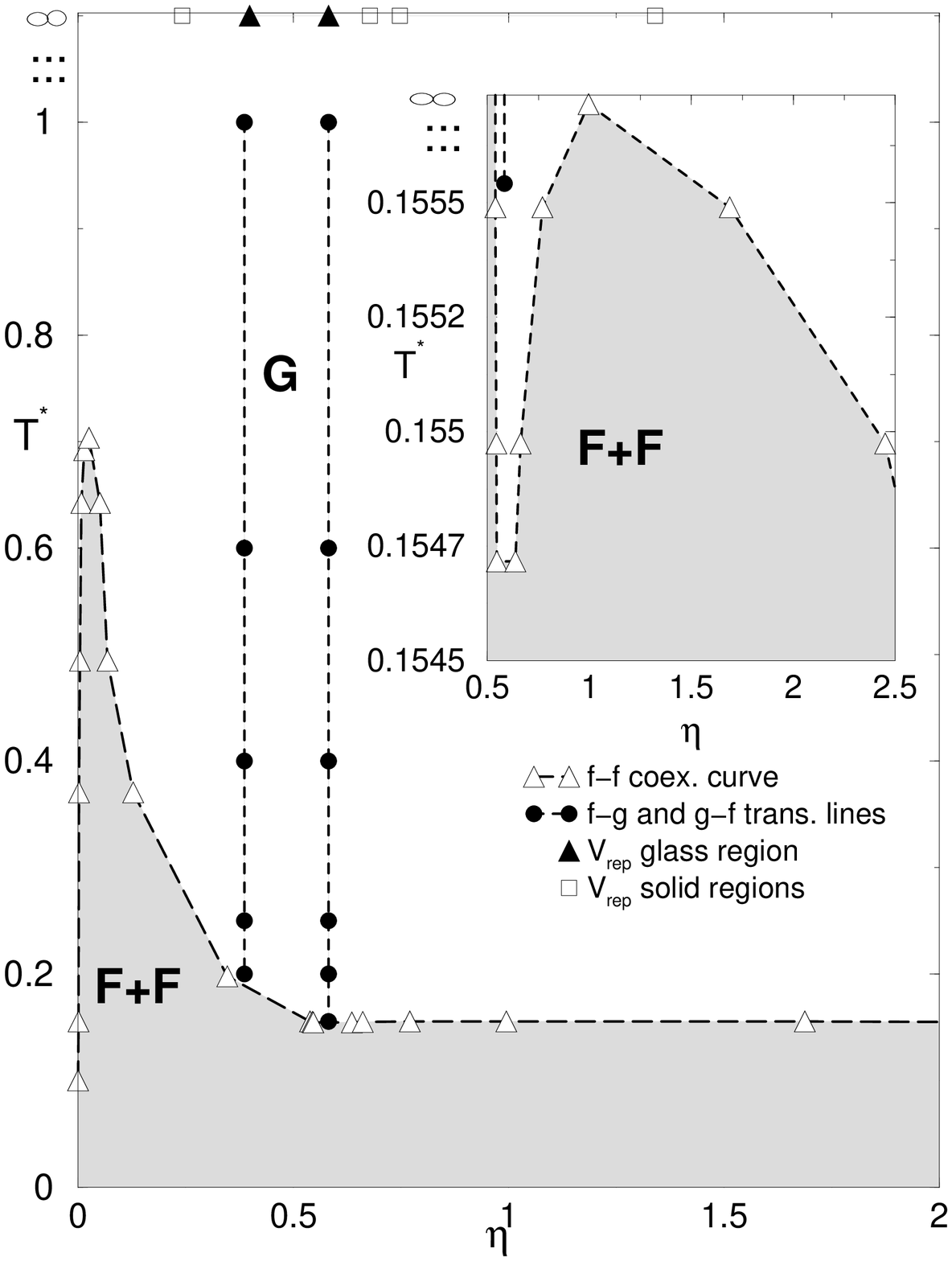,height=12cm}}
\caption{}

\label{fig6}
\end{figure}

\begin{list}
  {  \-{\bf FIG. }\arabic{nuovo}.} {\usecounter{nuovo}}

\item
$\beta V_{tot}(r)$ versus the interparticle separation $r/\sigma$
for two different temperature and $f=50$; $T^*=k_B T/C$.
We call $SP_1$ the interaction studied in ref.\cite{artmio} i.e. $A=2.1\sigma$,
$B=0.35\sigma$, $SP_2$ the interaction corresponding to $A=1.95\sigma$ and $B=0.21\sigma$, finally $SP_3$ corresponds to $A=1.875\sigma$ and
$B=0.155\sigma$. In figure we also show $V_{rep}$ corresponding to $f=50$.

\item
Athermal solvent: comparison between the main peak of the
 structure factors we obtained
with MHNC closure and MD simulation, with the results obtained by
Watzlawek {\it{et al.}} by means of MC simulation and RY closure;
${\eta}=0.60$, $f=32$.
($S(q)$ versus  $q \sigma$)


\item
Comparison between the radial distribution function (top) and the
 structure factor (bottom) we obtained
with the MHNC  closure (solid line) and MD simulation (opaque circles);
${\eta}=0.628$, $T^{*}=0.25$, $f=50$, $SP_2$.



\item
$SP_1$ fluid-fluid phase diagram (opaque triangles up) for a star polymer
solution calculated by means of mean field theory and
the ideal glass transition lines (filled circles)
 determined with MCT, which delimit the region where the system is a glass.
The inset shows
a magnification of the coexistence curve at higher density
\cite{artmio}.
On the top of the main figure we present some results about the purely
repulsive interaction, which corresponds to the limit of very high
temperature:
two successive
squares (from low to higher densities) delimit the regions where the system is solid at the
equilibrium (by Watzlawek {\it{et al.}}, MC simulation \cite{Phtd});
the filled triangles up represent the densities which delimit the
glass region.  Notice that the second critical point survives with
respect to the glass transition while the region around the second
critical point becomes metastable with respect to the freezing.
Moreover we can observe that the densities which delimit the glass
region are effectively the same with and without attractive
interaction.

\noindent
Lines are  simply a guides to the eye.

\item
MCT fluid-glass and glass-fluid lines (computed with MHNC)
for different $f$ values
and $A=1.95\sigma$, $B=0.21\sigma$ ($SP_2$): reduced temperature
$T^*=k_B T/C$ versus reduced packing fraction $\eta=\frac{\pi}{6}\rho \sigma^3$.  Opaque
symbols, circles, triangles, squares, stars and diamonds, correspond
respectively to $f=45,46,50,58,70$.
In the region between two line, fixed $f$,
the system is in a glass phase.
The equivalent
symbols
on the top of the
figure delimit the glass regions for $V_{rep}(r)$.
We recall that
in the limit of T very large we return to the simply repulsive system.

\noindent
Lines are  simply a guides to the eye.

\item
$SP_2$: nonergodicity parameter $f_q$ versus $q\sigma$ for stars with $f=50$,
$\eta=0.8$.
Notice the 
smaller width of this parameter for high temperatures.



\item
$SP_2$, $f=50$: magnification of  the
 structure factor (main peak) we obtained
with MD simulations  at different temperatures.
$a)$ ${\eta}=0.314$, $b)$  ${\eta}=0.628$; $f=50$.



\item
Reduced diffusion coefficient (MD) for different $f$ values and
$A=1.95\sigma$, $B=0.21\sigma$ ($SP_2$).  The symbols, plus,
circles, triangles, squares, correspond respectively to
$f=42,45,46,50$ (left panel $\eta=0.366$, right panel $\eta=0.628$).

\noindent
Lines are  simply a guides to the eye.

\item
$SP_2$: fluid-fluid phase diagram for a star polymer
solution calculated by means of mean field theory and
the ideal glass transition lines
 determined with MCT.
The legend is the same that in Fig.\ \ref{fig6}.
Notice that the second critical point
survives with respect to the glass transition
while the region around the second critical point
becomes metastable with respect to the freezing.
 Moreover we can observe that  decreasing the temperature
the densities which delimit the glass region move to higher values.
(in these cases
the effect is  small).

\noindent
Lines are  simply a guides to the eye.

\item
MCT fluid-glass  lines (computed with MHNC) for different $f$ values and
$A=1.875\sigma$,
$B=0.155\sigma$
($SP_3$): reduced temperature $T^*=k_B T/C$
versus reduced packing fraction $\eta=\rho\sigma^3$.
Filled   triangles, squares and diamonds
  correspond respectively to $f=46,50,70$.
The equivalent symbols on the top of the figure delimit the glass region corresponding to $V_{rep}(r)$.
We recall that in the limit of T very large we return to the simply repulsive system.

\item
Reduced diffusion coefficient (MD) for different $f$ values and $SP_3$.
The symbols, plus, circles, triangles, squares,  correspond respectively to $f=42,45,46,50$ (left panel $\eta=0.366$, right panel $\eta=0.628$).

\noindent
Lines are  simply a guides to the eye.

\item
$SP_3$, $f=46$: magnification of  the
 structure factor (main peak) we obtained
with MD simulations  at different temperatures.
$a)$ ${\eta}=0.55$,
$b)$ ${\eta}=0.51$.



\item
$SP_3$: fluid-fluid phase diagram  for a star polymer
solution calculated by means of mean field theory and
the ideal glass transition lines
 determined with MCT.
The legend is the same which in Fig.\ \ref{fig6}.

Notice that the second critical point
survives with respect to the glass transition
while the region around the second critical point
becomes metastable with respect to the freezing.
 Moreover we can observe that  decreasing the temperature
the densities which delimit the glass region move progressively
 to higher values.

\noindent
Lines are  simply a guides to the eye.

\item
Magnification of the low temperature MCT fluid-glass  lines (computed with MHNC) for different $f$ values, $SP_2,SP_3$.
The temperatures are rescaled with respect to the integrated intensity of the attractive contribution.

\noindent
Lines are  simply a guides to the eye.

\end{list}
\clearpage

\begin{figure}[ht]
\centerline{\psfig{figure=fig5.eps,height=9.5cm}}
\caption{}

\label{fig7}
\end{figure}

\begin{figure}[ht]
\centerline{\psfig{figure=fig6.eps,height=9cm}}
\caption{}

\label{fig8}
\end{figure}

\begin{figure}[ht]
\centerline{\psfig{figure=fig7.eps,height=10cm}}
\caption{}

\label{fig10bis}
\end{figure}

\begin{figure}[ht]
\centerline{\psfig{figure=fig8.eps,height=9.5cm}}
\caption{}

\label{fig9}
\end{figure}

\begin{figure}[ht]
\centerline{\psfig{figure=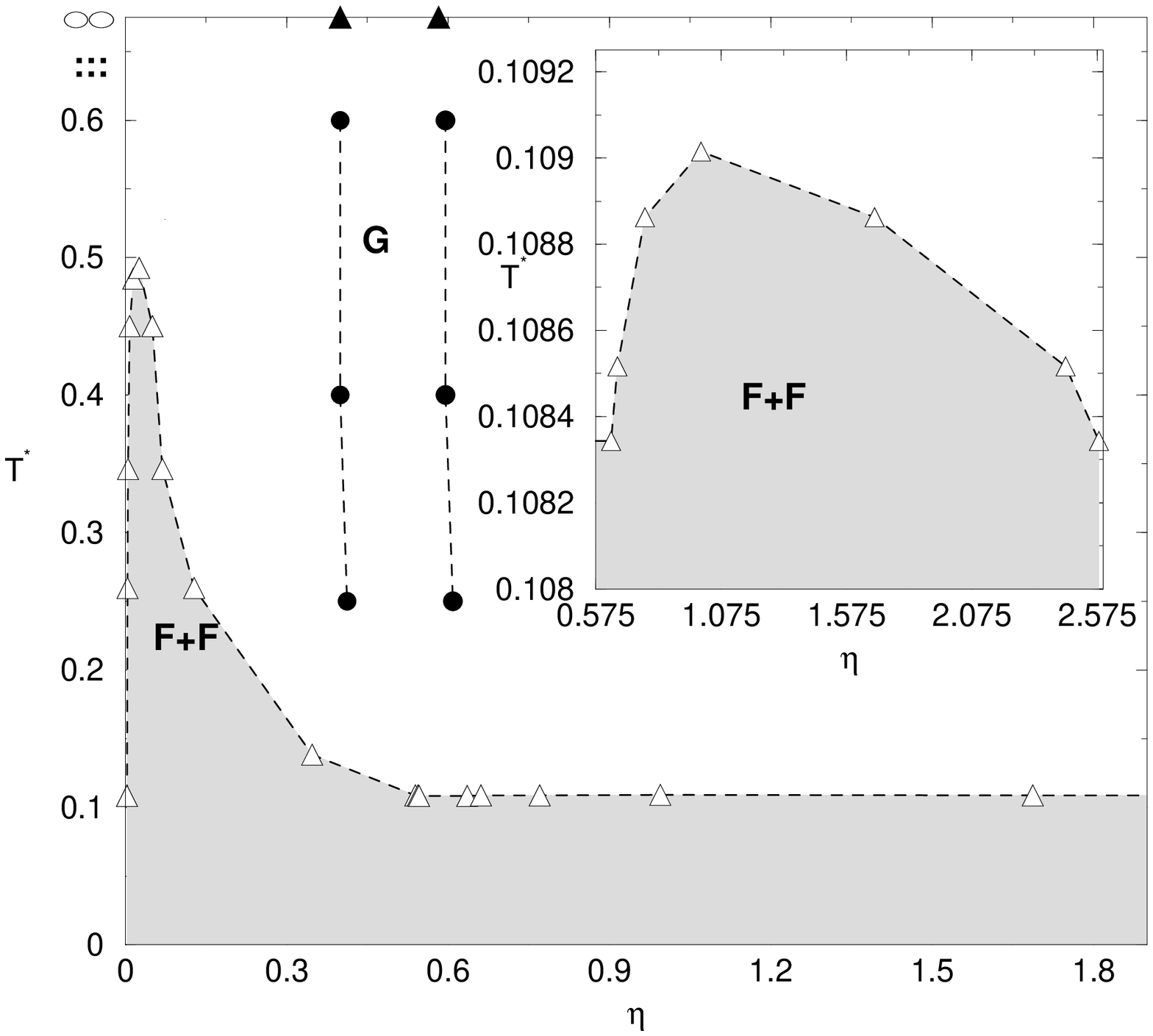,height=10.5cm}}
\caption{}

\label{fig13}
\end{figure}

\begin{figure}[ht]
\centerline{\psfig{figure=fig10.eps,height=8cm}}
\caption{}

\label{fig14}
\end{figure}

\begin{figure}[ht]
\centerline{\psfig{figure=fig11.eps,height=10cm}}
\caption{}

\label{fig15}
\end{figure}

\begin{figure}[ht]
\centerline{\psfig{figure=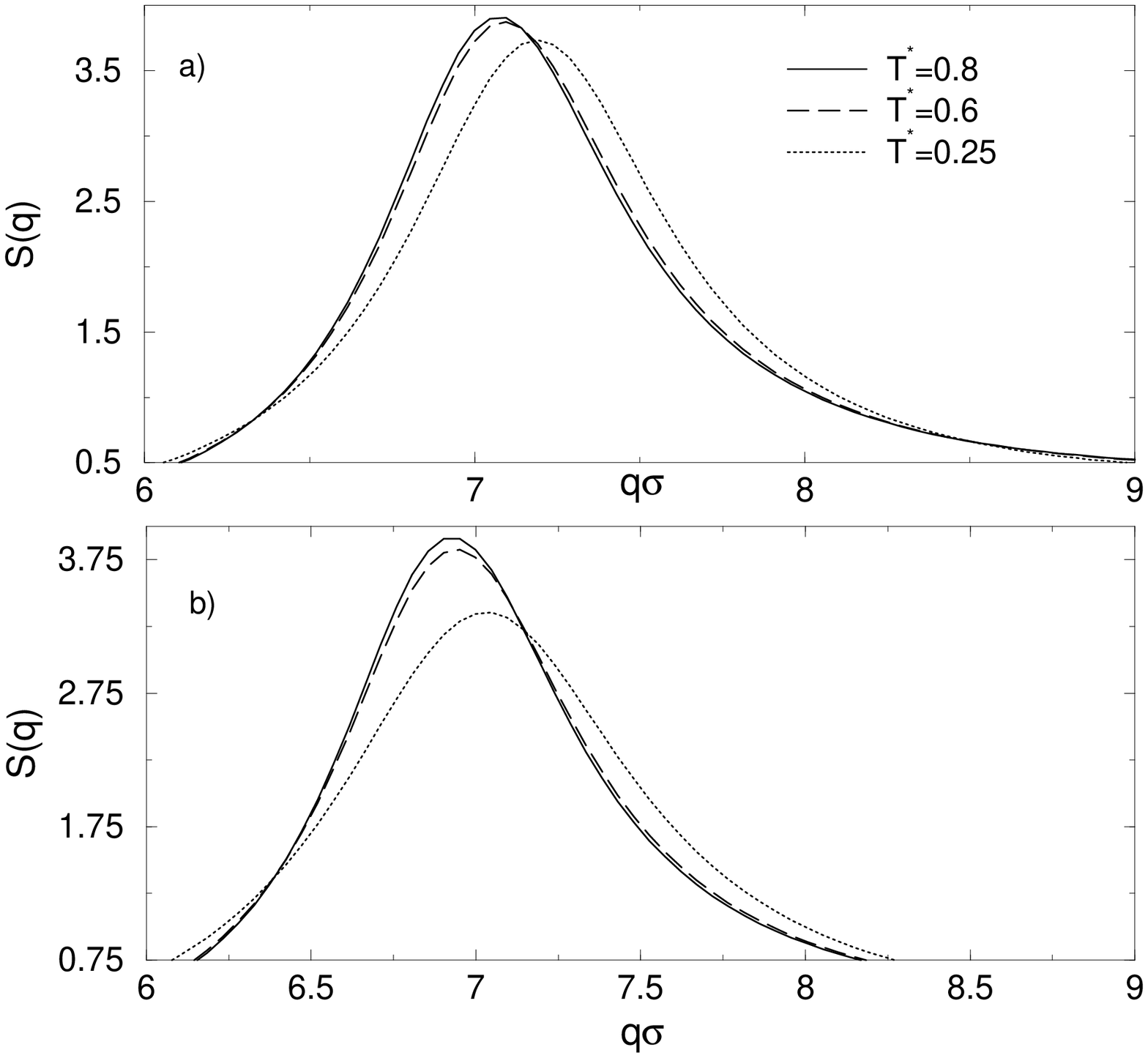,height=10cm}}
\caption{}

\label{fig16}
\end{figure}

\begin{figure}[ht]
\centerline{\psfig{figure=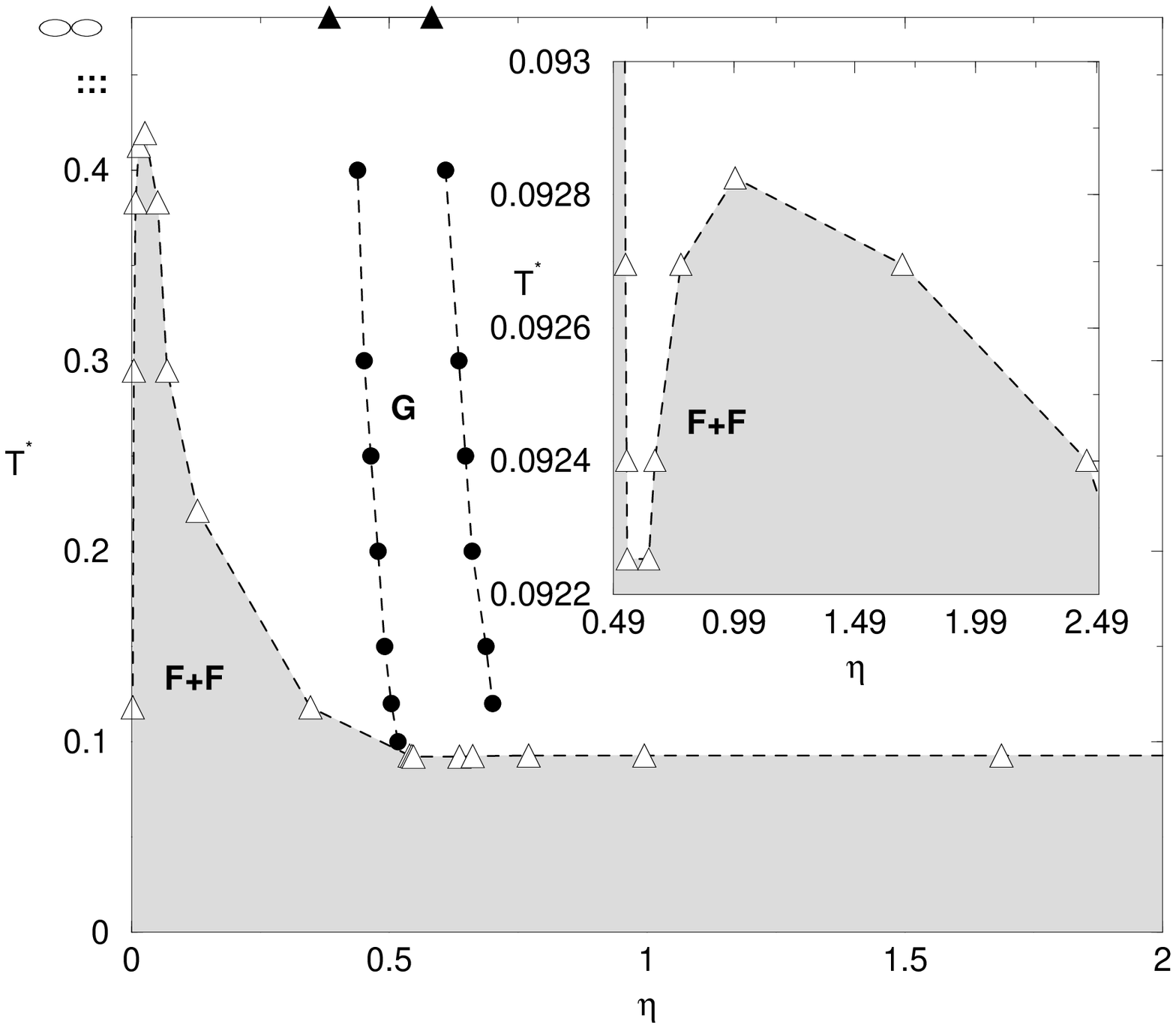,height=10cm}}
\caption{}

\label{fig18}
\end{figure}

\begin{figure}[ht]
\centerline{\psfig{figure=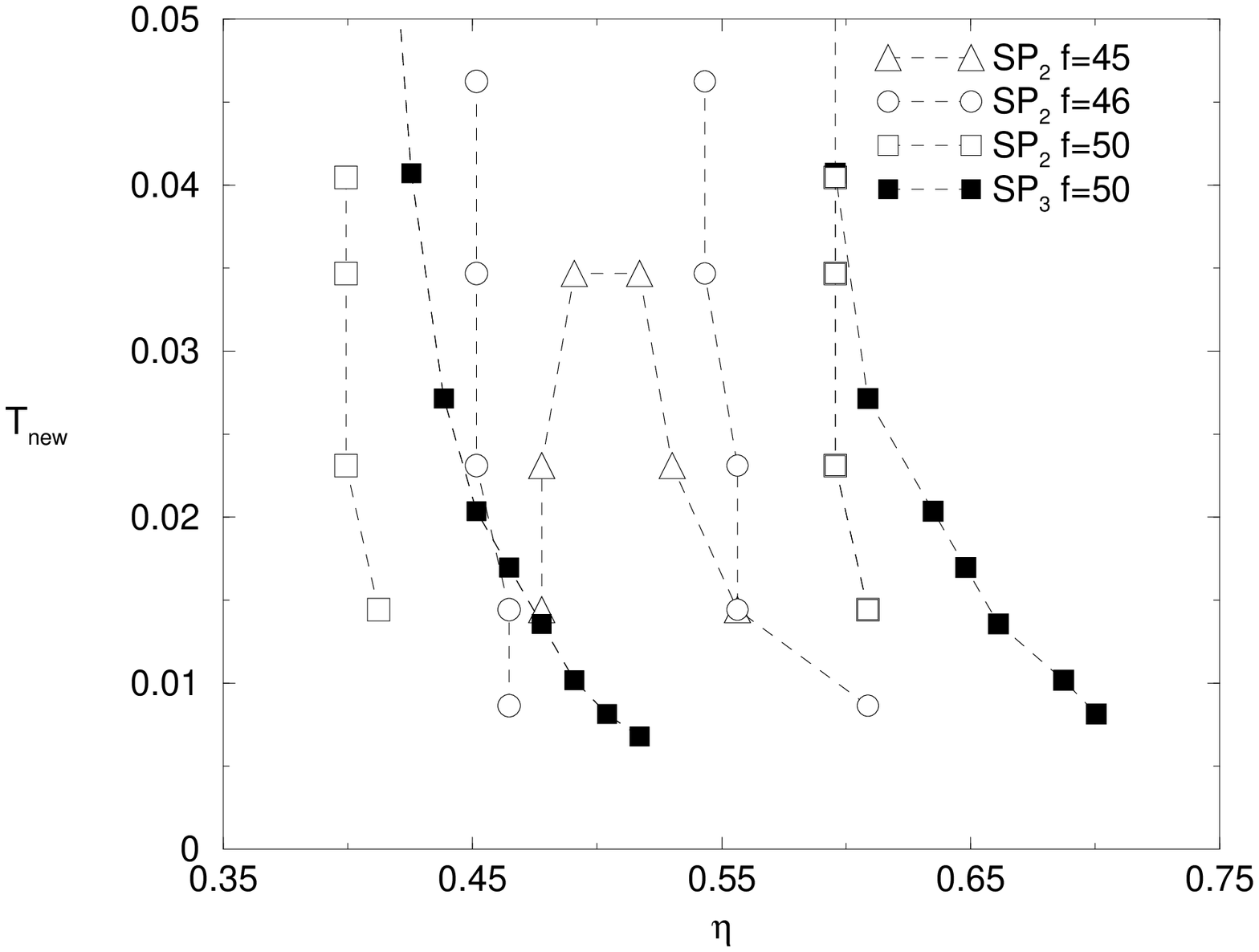,height=10cm}}
\caption{}

\label{fig19}
\end{figure}

\clearpage

\begin{thebibliography}{99}
\bibitem{footnote}
The upper size limit of a colloidal particle
is normally set at a radius of
$1$-$2\mu m$.
\end{thebibliography}
\end{document}